\documentclass[12pt]{JHEP3}
\usepackage{amssymb,amsmath,amsfonts}
\usepackage{cite}

\newcommand{\beqs}{\begin{equation*}}
\newcommand{\beq}{\begin{equation}}

\newcommand{\eeqs}{\end{equation*}}
\newcommand{\eeq}{\end{equation}}

\newcommand{\beqas}{\begin{eqnarray*}}
\newcommand{\beqa}{\begin{eqnarray}}

\newcommand{\eeqas}{\end{eqnarray*}}
\newcommand{\eeqa}{\end{eqnarray}}

\newcommand{\eq}[2]{\begin{equation} #1 \label{#2} \end{equation}}

\newcommand{\eps}{\varepsilon}

\newcommand{\de}{\delta}
\newcommand{\om}{\omega}

\newcommand{\la}{\lambda}
\newcommand{\si}{\sigma}

\newcommand{\La}{\Lambda}

\newcommand{\blist}{\begin{itemize}}

\newcommand{\elist}{\end{itemize}}

\providecommand{\href}[2]{#2}

\DeclareFontFamily{OT1}{rsfs}{}
\DeclareFontShape{OT1}{rsfs}{m}{n}{ <-7> rsfs5 <7-10> rsfs7 <10->rsfs10}{} 
\DeclareMathAlphabet{\mycal}{OT1}{rsfs}{m}{n}

\DeclareMathOperator{\extdm}{d}
\newcommand{\extd}{\extdm \!}

\newcommand{\tr}{{\rm tr}\;}

\title{Towards non-AdS holography in 3-dimensional higher spin gravity}

\author{Michael Gary, Daniel Grumiller and Radoslav Rashkov\footnote{On leave from the Department of Physics, Sofia University.}\\
           Institute for Theoretical Physics, 
           Vienna University of Technology,\\
           Wiedner Hauptstr. 8--10/136,
           A-1040 Vienna, Austria\\
           Email: \email{mgary@hep.itp.tuwien.ac.at, grumil@hep.itp.tuwien.ac.at, rash@hep.itp.tuwien.ac.at}}

\abstract{
We take the first steps towards non-AdS holography in higher spin gravity.
Namely, we propose a variational principle for generic 3-dimensional higher spin gravity that accommodates asymptotic backgrounds beyond AdS, like asymptotically Schr\"odinger, Lifshitz or warped AdS spacetimes.
As examples we study in some detail the four $sl(2)$ embeddings of spin-4 gravity and provide associated geometries, including an asymptotic Lifshitz black hole.
}


\preprint{TUW--12--01}

\begin{document}

\section{Introduction}

      One of the most striking and unexpected discoveries of the last two decades was the
AdS/CFT correspondence, according to which $N = 4$ supersymmetric Yang-Mills theory
in four dimensions is dual to type IIB superstring theory on AdS$_5\times S^5$ \cite{Maldacena:1997re}. 
Since then, the correspondence has been much generalized and has found many applications.
      Due to the dual nature of the AdS/CFT correspondence, it is rare to find an example
of a holographic duality where both sides of the correspondence are computable at the
same point in parameter space. After a decade of checks of the AdS/CFT correspondence,
in the last years there was impressive progress using techniques such as integrability \cite{Beisert:2010jr} and
supersymmetric localization \cite{Pestun:2007rz} that allowed computations for various regimes of the coupling.
Even within this framework the most interesting questions about quantum gravity remained
beyond the realization of AdS/CFT.
      However, if we can find the quantum gravity dual of an exactly soluble field theory (or vice versa),
then the AdS/CFT correspondence opens the possibility to understand quantum gravity
quantitatively.
      A natural playground for addressing these questions are two dimensional CFTs. The exact solubility of
these theories in generic regimes of parameter space makes them very attractive as toy
models for questions that are typically very hard for analogous field theories in higher
dimensions. Restricting to two dimensions on the field theory side implies that we should consider 
3-dimensional models on the gravity side. 

       One of the remarkable features of AdS spacetimes is the existence of interacting theories of massless particles 
with higher spins (greater than two) \cite{Fradkin:1986qy,Fradkin:1987ks,Vasiliev:1990en,Vasiliev:2003ev}. It has been suggested
\cite{Mikhailov:2002bp,Sezgin:2002rt} that these theories might be relevant for the description of certain sectors of large $N$
gauge theories. A striking conjecture was made by Klebanov and Polyakov \cite{Klebanov:2002ja} who proposed that a particular higher spin 
theory on AdS$_4$ might be exactly dual to sectors of
the free and interacting $O(N)$ vector model in $2 + 1$ dimensions at large $N$ realized under
different boundary conditions. This conjecture triggered an intensive study of the subject with
impressive achievements.
       In a series of papers the holographic study on both sides of the AdS/CFT correspondence has been
considered \cite{Giombi:2009wh,Giombi:2010vg,Koch:2010cy,Giombi:2011ya} --- three point functions, twistorial approach to holography with
higher spins, collective coordinates and critical $O(N)$.
       The authors of \cite{Douglas:2010rc} have shown that starting from free bosonic field theory one can
derive a dual description as a higher spin gravity in AdS space and reproduce
all correlation functions. A program of studying AdS$_3$ gravity duals of minimal model
CFTs was initiated in \cite{Gaberdiel:2010pz,Gaberdiel:2011zw}.

In three dimensions higher spin gravity theories are considerably simpler than in higher dimensions.
This is so because it is possible to truncate the otherwise infinite tower of higher spin fields
at arbitrary finite spin $n$ so that all fields have spin $s\leq n$ \cite{Aragone:1983sz}. Moreover, higher spin gravity in three dimensions 
can be formulated as a Chern--Simons gauge theory with gauge group $SL(n)\times SL(n)$ \cite{Blencowe:1988gj,Bergshoeff:1989ns}. 
(For $n=2$ the well-known Chern--Simons formulation of Einstein gravity with negative cosmological constant is recovered \cite{Achucarro:1987vz,Witten:1988hc,Carlip:1995zj}.)
With asymptotic AdS boundary conditions the asymptotic symmetry algebra is generated by two copies of $W_n$ algebras, which
can be shown using the Drinfeld--Sokolov reduction.
(For $n=2$ two copies of the Virasoro algebra are recovered.)
We refer to \cite{Bais:1990bs,Campoleoni:2010zq} and references therein for a more detailed summary of these constructions.

In many applications it is necessary to generalize the AdS/CFT correspondence to a gauge/gravity duality that does not involve spacetimes asymptoting to AdS, or asymptoting to AdS in a weaker way as compared to Brown--Henneaux \cite{Brown:1986nw,Henneaux:2010xg}.
Examples are null warped AdS spacetimes, which arise in proposed holographic duals of non-relativistic CFTs describing cold atoms \cite{Son:2008ye,Balasubramanian:2008dm}, Schr\"odinger spacetimes, which generalize null warped AdS by introducing an arbitrary scaling exponent \cite{Adams:2008wt}, Lifshitz spacetimes, which arise in gravity duals of Lifshitz-like fixed points \cite{Kachru:2008yh} and also have a scaling exponent parametrizing spacetime anisotropy, as well as the AdS/log CFT correspondence \cite{Grumiller:2008qz}, which requires a relaxation of the Brown--Henneaux boundary conditions \cite{Grumiller:2008es,Henneaux:2009pw,Maloney:2009ck}.

It is a priori not clear that higher spin gravity can accommodate such backgrounds.
There is, however, a precedent from which one may draw optimism, namely conformal Chern--Simons gravity \cite{Deser:1982vy,Deser:1982wh,Horne:1988jf}.
This theory bears resemblance to 3-dimensional higher spin gravity: it has no local physical degrees of freedom, it has a Chern--Simons formulation with gauge group bigger than $SL(2)\times SL(2)$, it has gauge symmetries that relate non-diffeomorphic metrics to each other, and the asymptotic symmetry algebra can be larger than two copies of the Virasoro algebra \cite{Afshar:2011yh,Afshar:2011qw}.
Its axisymmetric stationary solutions include AdS as well as AdS$_2\times\mathbb{R}$ spacetimes. 
Thus, at least for this precedent non-AdS backgrounds exist.

The main purpose of the present work is to show that higher spin gravity in three dimensions is suitable for the gauge/gravity duality beyond the AdS/CFT correspondence by constructing asymptotic non-AdS backgrounds.
We generalize the variational principle considered so far and show how to generate spacetimes that asymptote to AdS (with boundary conditions that can be weaker than Brown--Henneaux), to AdS$_2\times\mathbb{R}$, Schr\"odinger, Lifshitz or warped AdS spacetimes.
While our discussion is completely general and applies to arbitrary spin-$n$ theories, we focus for sake of specificity on the spin-4 case and provide examples for all four $sl(2)$-embeddings into $sl(4)$, including a Lifshitz black hole solution.

This paper is organized as follows. 
In section \ref{sec:2} we present our variational principle.
In section \ref{sec:3} we state how AdS backgrounds with non-Brown--Henneaux boundary conditions can be implemented.
In section \ref{sec:4} we demonstrate how to obtain specific non-AdS backgrounds, namely AdS$_2\times\mathbb{R}$, Schr\"odinger, Lifshitz and warped AdS spacetimes.
In section \ref{sec:5} we provide explicit spin-4 examples.
In section \ref{sec:6} we point to further applications and open issues.
In appendix \ref{app:A} we list suitable bases for all $sl(2)$-embeddings in the spin-4 case.

\section{Variational principle}\label{sec:2}

The bulk action of spin-$n$ gravity is the difference of two Chern--Simons actions with $sl(n)$ connections $A$ and $\bar A$ and associated field strengths $F$ and $\bar F$.
\eq{
S_{\rm bulk} = \frac{k}{4\pi}\,\int_{\cal M}\!\!\! \tr\big[CS(A)-CS(\bar A)\big]
}{eq:hs1}
The level $k$ is essentially the inverse Newton constant.
The 3-dimensional manifold $\cal M$ is supposed to have some smooth simply-connected boundary $\partial\cal M$.
The Chern--Simons 3-form is given by
\eq{
CS(A)= A\wedge \extd A + \tfrac23\,A\wedge A \wedge A
}{eq:hs2}
and similarly for $CS(\bar A)$.
We focus in the following just on one Chern--Simons copy, as the discussion for the other one is analogous.
The variation of the bulk action yields the bulk flatness conditions
\eq{
F = 0 
}{eq:hs3}
as well as boundary conditions
\eq{
\int_{\partial\cal M}\!\!\!\!\! \tr\big[ A\wedge\delta A \big] = 0\,.
}{eq:hs4}

As reviewed in \cite{Campoleoni:2010zq} (see \cite{Banados:1994tn,Banados:1998ta,Banados:1998gg,Carlip:2005zn} for earlier work) it is convenient to use Gaussian coordinates (with some ``radial'' coordinate $\rho$) and to parametrize the boundary $\partial\cal M$ in terms of ``light-cone'' variables $x^\pm$.
Employing a convenient partial gauge fixing with some group element $b(\rho)$
\eq{
A_\rho = b^{-1}(\rho)\partial_\rho b(\rho)
}{eq:hs5}
allows one to solve the flatness conditions $F_{\pm\rho}=0$ as follows.
\eq{
A_\pm =  b^{-1}(\rho) a_\pm (x^+,\,x^-) b(\rho)
}{eq:hs6} 
The Lie algebra valued functions $a_\pm(x^+,\,x^-)$ depend on the light-cone coordinates, but not on the radial coordinate $\rho$.
The flatness condition $F_{+-}=0$ is solved if 
\eq{
a_\pm=a_\pm(x^\pm) \qquad \textrm{and}\qquad [a_+,\,a_-] = 0\,.
}{eq:hs7}
While more general solutions of the flatness condition $F_{+-}=0$ are possible, in this work we exclusively consider the special case \eqref{eq:hs7}. 
In the parametrization above the boundary conditions \eqref{eq:hs4} translate into
\eq{
\int_{\partial\cal M}\!\!\!\!\! \tr\big[ A_+\wedge\delta A_- -  A_-\wedge\delta A_+ \big] = 0\,.
}{eq:hs8}
These boundary conditions are solved, for instance, by choosing one chirality of the gauge field to zero, $A_-=0$, at the boundary.
We note, however, that these are quite strong boundary conditions, in the sense that the field has not only vanishing variation, but also vanishing value at the boundary.

We intend to relax these boundary conditions such that $A_-$ is still fixed at the boundary, but not necessarily to zero.
In order to achieve this we add a boundary term to the bulk action \eqref{eq:hs1}.\footnote{%
This is the same kind of boundary term that appears for $U(1)$ gauge fields in AdS$_3$, see e.g.~\cite{Kraus:2006wn}.
}
\eq{
S=S_{\rm bulk} + \frac{k}{8\pi}\,\int_{\partial\cal M}\!\!\!\!\!\tr \big[A\wedge A + \bar A \wedge \bar A\big]
}{eq:hs9}
This addition changes the boundary conditions \eqref{eq:hs8} to
\eq{
\int_{\partial\cal M}\!\!\!\!\! \tr\big[ A_+\wedge\delta A_- \big] = 0\,.
}{eq:hs10}
Thus, we only have to require $\delta A_-=0$ at the boundary, but not necessarily $A_-=0$.
Similarly, we only have to require $\delta \bar A_+=0$ at the boundary, but not necessarily $\bar A_+=0$.
This is the variational principle we are going to use henceforth.

\section{AdS backgrounds beyond Brown--Henneaux}\label{sec:3}

To set the stage for non-AdS holography, in this section we review AdS boundary conditions for arbitrary $sl(2)$ embeddings into $sl(n)$, with arbitrary $n$.
We denote the $sl(2)$ generators by $L_0,\,L_\pm$ (see appendix \ref{app:A} for our conventions).

In the $sl(2)$ case the metric is constructed from the Chern--Simons gauge connections by first linearly combining them into vielbein and (dualized) spin connection.
\eq{
e = A-\bar A \qquad \om = A+\bar A
}{eq:hs11}
Then the metric follows from taking the trace over the vielbein bilinear form.
\eq{
g_{\mu\nu} = \frac12\,\tr\big[e_{\mu} e_{\nu}\big]
}{eq:hs12}

In the higher spin case the first linear combination \eqref{eq:hs11} leads to a quantity that could be called ``zuvielbein''.\footnote{DG is grateful to Guy Moore for inventing this expression during a seminar talk by Alejandra Castro at McGill University in September 2011.}
The metric is then defined as the trace over the zuvielbein bilinear form.\footnote{%
At least for the principal embedding the definition \eqref{eq:hs13} is the only natural one.
For other embeddings other definitions are possible, see the discussions in \cite{Campoleoni:2011hg,Castro:2011fm}.
Since we do not know whether there is a unique preferred definition of the metric for generic embeddings we shall stick to the definition \eqref{eq:hs13}. 
}
\eq{
g_{\mu\nu} := \frac12\,\tr\big[(A-\bar A)_{\mu} (A-\bar A)_{\nu}\big]
}{eq:hs13}
Due to the enhanced gauge symmetry two non-diffeomorphic metrics can be gauge equivalent.
Chern--Simons connections solving the flatness conditions discussed in section \ref{sec:2} are given by
\begin{subequations}
\label{eq:hs32}
\begin{align}
A&=b^{-1}(\rho)\big(a_+(x^+) b(\rho)\,\extd x^+ + a_-(x^-) b(\rho)\,\extd x^- +\partial_\rho b(\rho)\,\extd\rho\big) \label{eq:ha16} \\
\bar A&=\big(b(\rho)\bar a_+(x^+)\,\extd x^+ + b(\rho)\bar a_-(x^-)\,\extd x^- - \partial_\rho b(\rho)\,\extd\rho\big)\,b^{-1}(\rho) \label{eq:hs17}
\end{align}
\end{subequations}
with $[a_+(x^+),\,a_-(x^-)]=0=[\bar a_+(x^+),\,\bar a_-(x^-)]$.

So far in this paper the considerations were completely general.
Now we focus on asymptotic AdS boundary conditions.
For Poincar\'e patch AdS (with unit AdS radius) the connections are chosen as (see for instance \cite{Banados:1994tn,Coussaert:1995zp})
\begin{subequations}
\label{eq:hs31}
\begin{align}
 A_\rho^{\textrm{\tiny AdS}} &= L_0 &  \bar A^{\textrm{\tiny AdS}}_\rho &= -L_0 \\
 A_+^{\textrm{\tiny AdS}} &= e^{\rho} L_+  &  \bar A_+^{\textrm{\tiny AdS}} &= 0  \\
 A_-^{\textrm{\tiny AdS}} &= 0 & \bar A_-^{\textrm{\tiny AdS}} &= e^{\rho} L_- \,.
\end{align}
\end{subequations}
We call a class of connections asymptotically AdS if
\eq{
A - A^{\textrm{\tiny AdS}} = {\cal O}(e^{(1-\eps)\rho}) = \bar A - \bar A^{\textrm{\tiny AdS}} \qquad \textrm{for\;some\;}\eps > 0\,.
}{eq:hs26}
The boundary conditions \eqref{eq:hs26} are in general subject to further constraints from demanding finiteness, integrability and conservation of the canonical charges, a well-defined variational principle and the existence of two copies of the Virasoro algebra as part of the asymptotic symmetry algebra.

Let us consider first the Brown--Henneaux case, $\eps=1$ in the boundary conditions \eqref{eq:hs26}, and focus on configurations that exist in all embeddings of all higher spin gravity theories.
For this purpose the group element $b(\rho)$
\eq{
b(\rho) = e^{\rho L_0}
}{eq:hs14}
and the Lie-algebra valued functions $a_+$ and $\bar a_-$ are chosen in a standard way \cite{Coussaert:1995zp}.
\eq{
a_+ = L_+ + \ell_+(x^+)\,L_- \qquad \bar a_- = L_- + \ell_-(x^-)\,L_+
}{eq:hs15}
If we fix additionally $a_-=\bar a_+=0$ --- as required by the variational principle based upon the bulk action \eqref{eq:hs1} alone --- then the metric \eqref{eq:hs13} is asymptotically AdS ($a_0=\frac12\,\tr L_0^2$, $a_1=\frac12\,\tr\big(L_+L_-\big)$),
\eq{
\extd s^2 = a_0\,\extd\rho^2 + a_1\big(2\ell_+\,(\extd x^+)^2 + (e^{2\rho}+e^{-2\rho}\ell_+\ell_-)\,\extd x^+\extd x^- + 2\ell_-\,(\extd x^-)^2\big) 
}{eq:hs18}
and compatible with Brown--Henneaux boundary conditions \cite{Brown:1986nw}. 
The fact that the leading expression in the line-element \eqref{eq:hs18} grows like $e^{2\rho}$ and the subleading terms are order unity follows directly from trace properties and the Baker--Campbell--Hausdorff formula,
$b^{-1}(\rho) L_\pm b(\rho)=e^{\pm \rho} L_\pm$ and $b(\rho)L_\pm b^{-1}(\rho)=e^{\mp \rho} L_\pm$.

In order to find solutions that are not accommodated by the Brown--Henneaux boundary conditions we could either consider an Ansatz with fixed but nonzero $A_-$ and/or $\bar A_+$, which would require the boundary term in \eqref{eq:hs9};
or we replace the choice \eqref{eq:hs15} by an appropriate more general one.
As long as we are content with asymptotic AdS behavior we can implement the second option for most embeddings.
For instance, we may choose
\eq{
a_+ = L_+ + w_+(x^+)\,W_+ + \ell_+(x^+)\,L_- \qquad \bar a_- = L_- + w_-(x^-)\,W_- + \ell_-(x^-)\,L_+\,.
}{eq:hs19}
The generators $W_\pm$ are supposed to have the following properties:
\eq{
\tr\big(W_\pm L_\mp\big) = 0 \qquad \tr\big(W_+W_-\big)\neq 0 \qquad [W_\pm,\,L_0] = \pm h W_\pm\qquad h>0
}{eq:hs20}
The last condition excludes singlets.
The line-element constructed from \eqref{eq:hs13} reads 
\eq{
\extd s^2 = a_0\, \extd\rho^2 + 2a_1\, \ell(x^+) (\extd x^+)^2 + 2a_1\, \ell(x^-) (\extd x^-)^2 + \gamma_{+-}\,\extd x^+\extd x^-
}{eq:hs21}
with the Fefferman-Graham like expansion
\eq{
\gamma_{+-} = a_1\, e^{2\rho} + a_2\,w_+(x^+)w_-(x^-)\,e^{2h\rho} + a_1\, \ell_+(x^+)\ell_-(x^-)\,e^{-2\rho}\,.
}{eq:hs22}
The constants $a_i$ depend only on the values of various traces and are non-zero by assumption.
If $h=1$ then the expansion \eqref{eq:hs22} reduces essentially to the one of Brown--Henneaux, in the sense that the Fefferman--Graham expansion of the metric involves only integer powers in $e^{2\rho}$.
For any other (positive) $h$ the expansion \eqref{eq:hs22} goes beyond the one of Brown--Henneaux.

Some remarks are in order.
Whether or not the functions $w_\pm$ can be state-dependent or have to be fixed quantities depends on the specific theory, in particular on the $sl(2)$ weight $h$ of the generator $W_+$.
The choice \eqref{eq:hs19} is far from being generic and can easily be generalized to include more generators with positive or negative $sl(2)$ weights.
As we hinted above for some embeddings --- those where only singlets and triplets exist --- no essential extension of the Brown--Henneaux boundary conditions is possible.

Residual gauge transformations that respect the partial gauge fixing \eqref{eq:hs5} and the boundary condition $\de A_-|_{\partial\cal M}=0$ are generated by Lie-algebra valued functions $\Lambda$ that obey
\eq{
\de A_\rho = \partial_\rho \Lambda + [A_\rho,\,\Lambda] = 0 \qquad  \de A_-|_{\partial\cal M} = \partial_- \Lambda|_{\partial \cal M} + [A_-,\,\Lambda]|_{\partial\cal M} = 0\,.
}{eq:hs33}
The first condition \eqref{eq:hs33} is solved by $\La = b^{-1}(\rho)\la(x^+,\,x^-)b(\rho)$. 
In the case of an asymptotic boundary at $\rho\to\infty$ the second condition \eqref{eq:hs33} then restricts the Lie-algebra valued function $\la(x^+,\,x^-)$ to a sum over all generators $W_-^n$ that have a negative $sl(2)$ weight, $[W_-^n,\,L_0]=-h_n W_-^n$, with $h_n>0$.
\eq{
\la(x^+,\,x^-)=\sum_n w_n(x^+,\,x^-)\, W_-^n
}{eq:hs34}
For finite boundaries we obtain instead
\eq{
\la(x^+,\,x^-)=\sum_n w_n(x^+)\, C^n_{a_-}
}{eq:hs35}
where the sum extends over all generators $C^n_{a_-}$ that commute with $a_-$.
Analog considerations apply to residual gauge transformations of $\bar A$.
Finally, we note that our variational principle actually requires $\tr\big(A_+\wedge \de A_- - \bar A_-\wedge\de\bar A_+\big)_{\partial\cal M}=0$, so for asymptotic boundaries the conditions $\de A_-, \de\bar A_+|_{\partial M}\to 0$ are not always sufficient nor necessary.

\section{Non-AdS backgrounds}\label{sec:4}

In this section we demonstrate how the connections have to be chosen in order to obtain asymptotic geometries that differ from asymptotic AdS.
The main purpose of our discussion is to reveal that such choices are possible for certain classes of embeddings, but we refrain from performing an exhaustive scan.
We start with AdS$_2\times\mathbb{R}$ in section \ref{sec:4.3}, follow up with  Schr\"o\-din\-ger spacetimes in section \ref{sec:4.2}, Lif\-shitz spacetimes in section \ref{sec:4.4}, and then consider warped AdS in section \ref{sec:4.1}.
The discussion will be general and focused on the leading asymptotic behavior.

\subsection{AdS$_2\times\mathbb{R}$ background}\label{sec:4.3}

Surprisingly little structure is needed to construct direct product spaces of maximally symmetric spacetimes.
We focus here on AdS$_2\times\mathbb{R}$ and prove constructively that any $sl(2)$ embedding of spin-$n$ gravity that has at least one singlet $S$ with $\tr S^2\neq 0$ allows such backgrounds.
Consider the connections ($L_n$ are the $sl(2)$ generators)
\begin{subequations}
\label{eq:hs42}
\begin{align}
 A_\rho &= L_0 & \bar A_\rho &= -L_0 \\
 A_1 &= a_1\,e^{\rho} L_+ &  \bar A_1 &= e^{\rho} L_-  \\
 A_2 &= 0 & \bar A_2 &= S 
\end{align}
\end{subequations}
with some constant $a_1\neq 0$.
We have relabeled the coordinates $x^\pm$ as $x^1, x^2$ to emphasize that they no longer refer to light cone components.
The choice \eqref{eq:hs42} is compatible with our variational principle in section \ref{sec:2} and solves all flatness conditions, since $[S,\,L_-]=0$ by assumption.
One can add further generators to the connections to generate subleading terms, but for sake of clarity we focus here just on the essential terms.
The non-vanishing metric components \eqref{eq:hs13} constructed from the connections \eqref{eq:hs42} read
\begin{subequations}
\label{eq:hs43} 
\begin{align}
g_{\rho\rho} &= 2\,\tr L_0^2 \\
g_{11} &= -a_1\,\tr\big(L_+L_-\big)\, e^{2\rho} \\      
g_{22} &= \frac12\,\tr S^2  \,.                  
\end{align}
\end{subequations}
Depending on the sign of $a_1$ the metric \eqref{eq:hs43} is locally and asymptotically AdS$_2\times\mathbb{R}$ or $\mathbb{H}_2\times\mathbb{R}$ ($\mathbb{H}_2$ is the Lobachevsky plane).

\subsection{Schr\"odinger background}\label{sec:4.2}

Asymptotic Schr\"odinger spacetimes \cite{Son:2008ye,Balasubramanian:2008dm,Adams:2008wt}
\eq{
\extd s^2 = \ell^2\,\Big[\frac{\extd r^2 \pm 2\extd t\extd\xi}{r^2} - \frac{\extd t^2}{r^{2z}}\Big]
}{eq:hs44}
are characterized by a scaling exponent $z\in\mathbb{R}$.
If $z=1$ ($z=2$) asymptotic (null warped) AdS is recovered.
Otherwise genuine Schr\"odinger spacetimes are obtained.

The simplest construction of such spacetimes in higher spin gravity requires only two generators $W_\pm$ in addition to the $sl(2)$ generators $L_n$, with the following properties:
\eq{
[W_\pm,\, L_0] = \pm z\,W_\pm\qquad [W_-,\,L_-] = 0\qquad \tr\big(W_+W_-\big) \neq 0\quad \tr\big(W_n L_m\big) = 0
}{eq:hs46}
Consider the connections
\begin{subequations}
\label{eq:hs45}
\begin{align}
 A_\rho &= L_0 & \bar A_\rho &= -L_0 \\
 A_t &= a_1 e^{\rho} L_+ + a_2 e^{\rho z} W_+  &  \bar A_t &=  e^{\rho z} W_-  \\
 A_\xi &= 0 & \bar A_\xi &= e^{\rho} L_- 
\end{align}
\end{subequations}
where we omit possible subleading terms for sake of conciseness.
Note that $\bar A_t\neq 0$, so we cannot use the variational principle based upon the bulk action \eqref{eq:hs1} alone.
Instead, we use the full action \eqref{eq:hs9} which allows $\bar A_t\neq 0$ at the boundary.
The connections \eqref{eq:hs45} are not only consistent with our variational principle, but also solve all flatness conditions and are thus valid solutions of any higher spin gravity theory that allows for generators with the properties \eqref{eq:hs46}.

With suitable choices for the constants $a_i$ in \eqref{eq:hs45} and the coordinate transformation $r = e^{-\rho z}$ the metric \eqref{eq:hs13} constructed from the connections \eqref{eq:hs45} leads precisely to the Schr\"odinger line element \eqref{eq:hs44}.
The same set of generators leads to an asymptotic Schr\"odinger spacetime with scaling exponent $1/z$, upon exchanging the components $\bar A_t$ and $\bar A_\xi$.

A construction similar to the one above works if there are two pairs of generators $W_\pm^{[1,2]}$ with the properties
\eq{
[W_\pm^{[i]},\, L_0] = \pm h^{[i]}\,W_\pm^{[i]}\qquad [W_-^{[i]},\,W_-^{[j]}] = 0\qquad \tr\big(W_+^{[i]}W_-^{[j]}\big) = t_i \de_{i,j}\quad t_i \neq 0 \,.
}{eq:hs50}
The weights $h^{[i]}$ are non-zero by assumption.
The scaling exponent is then given by
\eq{
z = \frac{h^{[1]}}{h^{[2]}} \qquad \textrm{or} \qquad z = \frac{h^{[2]}}{h^{[1]}} 
}{eq:hs51}
depending on whether $W_\pm^{[1]}$ or $W_\pm^{[2]}$ replace the generators $W_\pm$ in the connections \eqref{eq:hs45} (the other pair then replaces $L_\pm$).
More generally, if there are several pairs of generators with the properties \eqref{eq:hs50} then a larger set of scaling exponents \eqref{eq:hs51} is possible.

We note finally that subleading terms can be added to the Schr\"odinger connections \eqref{eq:hs45}, provided these terms do not spoil the flatness conditions \eqref{eq:hs7}. 

\subsection{Lifshitz background}\label{sec:4.4}

Asymptotic Lifshitz spacetimes \cite{Kachru:2008yh}
\eq{
\extd s^2 = \ell^2\,\Big[\frac{\extd r^2 + \extd x^2}{r^2} - \frac{\extd t^2}{r^{2z}}\Big]
}{eq:hs47}
are characterized by a scaling exponent $z\in\mathbb{R}$ as well.
The construction of such spacetimes in higher spin gravity is possible too.
Consider the connections
\begin{subequations}
\label{eq:hs48}
\begin{align}
 A_\rho &= L_0 & \bar A_\rho &= -L_0 \\
 A_t &= a_1\,e^{\rho z} W_+  &  \bar A_t &= e^{\rho z} W_- \\
 A_x &= e^{\rho} L_+ & \bar A_x &= a_2\, e^{\rho} L_- 
\end{align}
\end{subequations}
where again we omit possible subleading terms.
Again the constants $a_i$ are fixed suitably. 
We require the properties
\eq{
[W_\pm,\,L_0]=\pm z\,W_\pm\qquad [W_\pm,\,L_\pm]= 0\qquad \tr\big(W_+ W_-\big)\neq 0\quad \tr\big(W_n L_m\big) = 0
}{eq:hs49}
to ensure that the connections \eqref{eq:hs48} solve the flatness conditions and have the right scaling and trace properties.
With the coordinate transformation $r=e^{-\rho z}$  the metric \eqref{eq:hs13} constructed from the connections \eqref{eq:hs48} leads precisely to the Lifshitz line element \eqref{eq:hs47}.

Remarks analogous to the ones in the last two paragraphs of the Schr\"odinger section \ref{sec:4.2} apply also to the Lifshitz case.

\subsection{Warped AdS background}\label{sec:4.1}

In the constructions so far we did not need a lot of structure: in addition to the $sl(2)$ generators a singlet was sufficient for the AdS$_2\times\mathbb{R}$ case and a doublet (or another suitable pair of generators) for Schr\"odinger or Lifshitz cases.
We demonstrate now that warped AdS is only slightly more complicated than the previous cases.

Spacelike warped AdS is given by the line element \cite{Nutku:1993eb} (see \cite{Bengtsson:2005zj,Anninos:2008fx} for details and further references) 
\eq{
\extd s^2 = \frac{\ell^2}{\nu^2+3}\,\big(\extd\rho^2-\cosh^2\!\rho\,\extd t^2 + \frac{4\nu^2}{\nu^2+3}\,(\extd x + \sinh\rho\,\extd t)^2\big)\,.
}{eq:hs28}
Here $\ell$ is some length scale and $\nu$ is the warping parameter.
If $\nu^2>1$ ($\nu^2<1$) then we have stretched (squashed) AdS.
In the large $\rho$ expansion and with suitable rescalings\footnote{The rescalings are well-defined only for $\nu^2\neq 1$. However, after performing the rescalings we can consistently take the limit $\nu^2\to 1$ in order to obtain null warped AdS.} of the coordinates $t$ and $x$ the metric components read
\begin{subequations}
\label{eq:hs29} 
\begin{align}
g_{\rho\rho} &= \frac{\ell^2}{\nu^2+3} \\
g_{tt} &= \sigma\,e^{2\rho} + {\cal O}(1)\\                         
g_{tx} &= e^\rho + {\cal O}(e^{-\rho}) \\
g_{xx} &= \frac{3|\nu^2-1|}{4\nu^2} 
\end{align}
\end{subequations}
where $\si = \textrm{sign}\,(\nu^2-1)$.
In the limit $\nu^2\to 1$ the metric \eqref{eq:hs29} asymptotes to null warped AdS, which has curious properties \cite{Anninos:2010pm,Guica:2011ia}.
Our goal is to obtain a line element in some higher spin gravity theory whose asymptotics coincides with \eqref{eq:hs29}.

Consider the connections 
\begin{subequations}
\label{eq:hs27}
\begin{align}
 A_\rho &= L_0 & \bar A_\rho &= -L_0 \\
 A_t &= a_1 e^{\rho} L_+ + a_2 e^{\rho/2} W_+ + \ell(t)\,e^{-\rho} L_- &  \bar A_t &= e^{\rho} L_-  \\
 A_x &= 0 & \bar A_x &= e^{\rho/2} W_- + \mu\,S 
\end{align}
\end{subequations}
Here $\mu$ is some real parameter that defines the amount of warping, as we explain below, while the parameters $a_i$ will be chosen conveniently.
In addition to the $sl(2)$ generators $L_n$ we need three more generators $W_\pm$ and $S$ with the properties
\eq{
[W_\pm,\,L_0] = \pm \frac12 W_\pm\qquad[W_-,\,L_-]=[S,\,L_-] = 0
}{eq:hs23}
and the non-vanishing traces
\eq{
\tr\big(W_+W_-\big)=a_3\,\tr\big(L_+L_-\big)\qquad \tr S^2=a_4\,\tr\big(L_+L_-\big)\,.
}{eq:hs24}

The non-vanishing metric components read
\begin{subequations}
\label{eq:hs30} 
\begin{align}
g_{\rho\rho} &= 2\,\tr L_0^2 \\
g_{tt} &= -a_1\tr\big(L_+L_-\big)\,\big(e^{2\rho} - \ell(t) \big)\\      
g_{tx} &= -\frac12\,\tr\big(L_+L_-\big)\, a_2 a_3 e^{\rho} \\
g_{xx} &= \mu^2\,\tr\big(L_+L_-\big)\,\frac{a_4}{2}                    
\end{align}
\end{subequations}
If we choose the length scale $\ell$ appropriately in \eqref{eq:hs29} and fix $a_1$ and $a_2$ such that $-\frac12\,\tr\big(L_+L_-\big)\, a_2 a_3 = 1$ and $-a_1 \tr\big(L_+L_-\big)=\si$ then we recover the warped AdS result \eqref{eq:hs29} with warping parameter
\eq{
\big|1-\frac{1}{\nu^2}\big| = \mu^2\,\tr\big(L_+L_-\big)\,\frac{2a_4}{3} \geq 0\,.
}{eq:hs25}
Null warped AdS, $\nu^2=1$, is obtained for $\mu=0$.
The inequality in \eqref{eq:hs25} is satisfied since $\tr S^2>0$.

From the discussion above it is clear that the choice \eqref{eq:hs27} leads to a spacetime that asymptotes to spacelike warped AdS.
Interestingly, if we choose $\ell(t)=0$ then spacetime is not only asymptotically but also locally warped AdS, in the sense that all polynomial curvature invariants are constant and coincide with the ones of warped AdS.

We call a spacetime asymptotically spacelike warped AdS if the connections behave to leading order in a large $\rho$ expansion as in \eqref{eq:hs27}. 
As in the asymptotic AdS case the allowed behavior of the subleading terms is again subject to constraints from demanding finiteness, integrability and conservation of the canonical charges.
In topologically massive gravity \cite{Deser:1982vy,Deser:1982wh} these constraints (and additional ones) are accounted for by the Comp{\`e}re--Detournay boundary conditions \cite{Compere:2009zj}.

The discussion above was restricted to spacelike warped AdS, with null warped AdS as a possible limiting case.
It is straightforward to repeat the discussion for timelike warped AdS.
It turns out that the convexity condition analog to \eqref{eq:hs25} cannot be fulfilled.
Thus, timelike warped AdS is not accessible starting with the Ansatz \eqref{eq:hs27}.
However, in cases where more singlets exist it is possible to obtain timelike warped AdS.
Namely, replacing $\mu S$ in \eqref{eq:hs27} by $\mu_+ S^{[+]} + \mu_- S^{[-]}$ does the job, provided that $\tr\big(S^{[+]}S^{[-]}\big)\neq 0$ and the sign of $\mu_+\mu_-$ is chosen appropriately.
[The simplest example where $\tr\big(S^{[+]}S^{[-]}\big)\neq 0$ is the 2-1-1 embedding in spin-4 gravity, discussed in section \ref{sec:5.4} below.]

Finally, we mention that a construction very similar to the one above is possible if there are generators $W_\pm$ whose $sl(2)$ weight is $\pm 2$.
In that case the r\^ole of $L_\pm$ and $W_\pm$ in \eqref{eq:hs27} is essentially interchanged, and the AdS radius is rescaled by a factor of $2$.
More generally, asymptotically warped AdS can emerge as a background solution whenever two generators exist whose $sl(2)$ weights differ by a factor of $2$, provided the traces analog to \eqref{eq:hs24} do not vanish.

\section{Spin-4 examples}\label{sec:5}

In this section we provide explicit examples that realize various features discussed in the general sections \ref{sec:2}-\ref{sec:4} above.
The connections in our examples belong to $sl(4)$.
We follow the conventions of appendix \ref{app:A} and use the bases provided therein.

\subsection{2-2 embedding}\label{sec:5.2}

We start by considering the 2-2 embedding.
Using the generators defined in appendix \ref{app:A.2} it is easy to show that all of them have integer $sl(2)$ weights, with eigenvalues $\pm 1$ or $0$.
Choosing again $b(\rho)=e^{\rho L_0}$ in \eqref{eq:hs5} we consider the connections
\begin{subequations}
\label{eq:hs37}
\begin{align}
 A_\rho &= L_0 & \bar A_\rho &= -L_0 \\
 A_1 &= e^{\rho}\,W_+ + f_1\, W_0 + e^{-\rho}\,h_1\, W_- &  \bar A_1 &= 0 \\
 A_2 &= 0 & \bar A_2 &= e^\rho\, W_- + f_2\, W_0 + e^{-\rho}\,h_2\,W_+
\end{align}
\end{subequations}
Here $W_n$ stands schematically for all possible generators with $sl(2)$ weight $n$ and $f_{1,2} = f_{1,2}(x^{1,2})$, $h_{1,2}=h_{1,2}(x^{1,2})$.
Note that the Ansatz \eqref{eq:hs37} is generic for solutions obeying the boundary conditions $A_2=0=\bar A_1$ and the gauge fixing condition \eqref{eq:hs5}.
Since $\tr W_\pm^2=0=\tr W_\pm W_0$ the line element is asymptotic AdS and compatible with Brown--Henneaux boundary conditions.
It is not possible to construct asymptotic AdS solutions that violate the Brown--Henneaux boundary conditions, nor generic warped AdS solutions, nor Schr\"odinger/Lifshitz solutions for this embedding.
Constructing an AdS$_2\times\mathbb{R}$ (or $\mathbb{H}_2\times\mathbb{R}$) background is possible using the connections \eqref{eq:hs42} with $S=S^{[0]}$.

\subsection{3-1 embedding}\label{sec:5.3}

The 3-1 embedding is somewhat similar to the non-principal embedding in spin-3 gravity, studied for instance in \cite{Ammon:2011nk}, since it has exactly one singlet, one higher-spin multiplet (in the present case a quintet, in the spin-3 case a triplet) and a couple of multiplets with lower non-vanishing spin (here three triplets, in the spin-3 case two doublets).
As we show now the 3-1 embedding allows asymptotically AdS solutions beyond Brown--Henneaux.
Using the generators defined in appendix \ref{app:A.3} and $b=e^{\rho L_0/2}$ we consider the connections
\begin{subequations}
\label{eq:hs36}
\begin{align}
 A_\rho &= \frac12\,L_0 & \bar A_\rho &= -\frac12\,L_0 \\
 A_+ &= e^{\rho}\, W_2 + \ell_+(x^+)\,e^{\rho/2}\, L_+ + \dots &  \bar A_+ &= 0  \\
 A_- &= 0 & \bar A_- &= e^{\rho}\,W_{-2} + \ell_-(x^-)\,e^{\rho/2}\, L_- + \dots
\end{align}
\end{subequations}
The ellipsis refers to subleading terms.
Inserting the connections \eqref{eq:hs36} into the definition for the metric \eqref{eq:hs13} yields
\eq{
\extd s^2=\extd\rho^2 - \big(e^{2\rho} - 4\ell_+(x^+)\ell_-(x^-) e^\rho + \dots \big)\extd x^+\extd x^-\,.
}{eq:hs38}
If $\ell_\pm \neq 0$ then the line element \eqref{eq:hs38} asymptotes to AdS but violates the Brown--Henneaux boundary conditions.
While no genuine Schr\"odinger/Lifshitz spacetimes can be constructed here, generic warped AdS spacetimes exist for this embedding, since the $sl(2)$ weight of $W_2$ is $2$ and due to the existence of a singlet. 
We give an explicit warped AdS example in the next subsection.

\subsection{2-1-1 embedding}\label{sec:5.4}

The 2-1-1 embedding allows the construction of warped AdS solutions precisely along the lines of section \ref{sec:4.1}. 
Using the generators defined in appendix \ref{app:A.4} and $b(\rho)=e^{\rho L_0}$ we consider the connections
\begin{subequations}
\label{eq:hs39}
\begin{align}
 A_\rho &= L_0 & \bar A_\rho &= -L_0 \\
 A_t &= \sigma\,e^{\rho}\, L_+ + e^{\rho/2}\, G_+^{[1]} &  \bar A_t &= e^{\rho}\,L_-  \\
 A_x &= 0 & \bar A_x &= e^{\rho/2}\, G_-^{[3]} + \mu\,S 
\end{align}
\end{subequations}
We could add to $A_t$ further doublet generators $G_+^{[i]}$ as well as singlet generators $S,\,S^{[n]}$ and generators with negative $sl(2)$ weight. 
Moreover, we could add to $\bar A_x$ further doublet generators $G_-^{[i]}$ and the other singlet generators $S^{[n]}$.
To reduce clutter we focus instead on the simpler connections \eqref{eq:hs39}, as they are sufficient to produce an asymptotic (and locally) warped AdS background.
Inserting the connections \eqref{eq:hs39} into the definition for the metric \eqref{eq:hs13} yields
\eq{
\extd s^2=\extd\rho^2 + \sigma\,e^{2\rho}\,\extd t^2 + 2\,e^{\rho}\,\extd t\extd x + \frac{3(\nu^2-1)}{\nu^2}\, \extd x^2\,.
}{eq:hs40}
If $\sigma=\textrm{sign}(\nu^2-1)=1$ ($\sigma=-1$) the line element \eqref{eq:hs40} describes locally spacelike stretched (squashed) AdS \eqref{eq:hs29} with $\ell=\sqrt{\nu^2+3}$  and warping parameter $\nu$.
The latter is related to the constant $\mu$ in the connection component $\bar A_x$ \eqref{eq:hs39} through
\eq{
\big|1-\frac{1}{\nu^2}\big| = \frac23\,\mu^2 \,.
}{eq:hs41}
Setting to zero $\mu$ yields locally null warped AdS.

\subsection{Principal embedding}\label{sec:5.1}

The principal embedding has pairs of generators with non-vanishing trace with $sl(2)$ weights $1,2$ and $3$.
This implies that we can construct Schr\"odinger and Lifshitz spacetimes with the following list of scaling exponents.
\eq{
z \in \{\frac13,\,\frac12,\,\frac23,\,\frac32,\,2,\,3\}
}{eq:hs52}
As an example we pick connections leading to a $z=3$ Lifshitz spacetime.
\begin{subequations}
\label{eq:hs53}
\begin{align}
 A_\rho &= L_0 & \bar A_\rho &= -L_0 \\
 A_t &= a_1\, e^{3\rho} U_3 &  \bar A_t &= e^{3\rho}\,U_{-3}  \\
 A_x &= e^\rho\,L_+ & \bar A_x &= a_2\,e^\rho\,L_- 
\end{align}
\end{subequations}
This Ansatz together with appropriate choices for the constants $a_1$, $a_2$ lead to the line element ($r=e^{-\rho}=1/\hat r$)
\eq{
\extd s^2 = \ell^2\,\Big[\frac{\extd r^2 + \extd x^2}{r^2} - \frac{\extd t^2}{r^6}\Big] = \ell^2\,\Big[\frac{\extd\hat r^2}{\hat r^2} + \hat r^2\,\extd x^2 -\hat r^6\,\extd t^2 \Big]
}{eq:hs54}
with $\ell=\sqrt{10}$.

An interesting class of asymptotic Lifshitz solutions that contain subleading terms is given by the connections
\begin{subequations}
\label{eq:hs55}
\begin{align}
 A_\rho &= L_0 & \bar A_\rho &= -L_0 \\
 A_t &= \sum_{n=-3}^3 a_n\, e^{n\rho} U_n & \bar A_t &= \sum_{n=-3}^3 \bar a_n\, e^{-n\rho} U_n  \\
 A_x &= \sum_{n=-1}^1 f_n\, e^{n\rho} L_n & \bar A_x &= \sum_{n=-1}^1 \bar f_n\, e^{-n\rho} L_n
\end{align}
\end{subequations}
with the conditions
\begin{subequations}
 \label{eq:hs56}
\begin{align}
 a_3 &= \frac{a_1 f_1^2}{3 (f_0^2 + f_1 f_{-1})} & a_2 &= \frac{a_1 f_0 f_1}{f_0^2 + f_1 f_{-1}} & a_0 &= \frac{a_1 f_0 (f_0^2 + 6 f_1 f_{-1})}{3 f_1 (f_0^2 + f_1 f_{-1})} \\
 a_{-3} &= \frac{a_1 f_{-1}^3}{3 f_1 (f_0^2 + f_1 f_{-1})} & a_{-2} &= \frac{a_1 f_0 f_{-1}^2}{f_1 (f_0^2 + f_1 f_{-1})} & a_{-1} &= \frac{a_1 f_{-1}}{f_1} 
\end{align}
\end{subequations}
and the same kind of conditions for the coefficients $\bar a_n$.
It is not trivial, but true, that the connection $A$ in \eqref{eq:hs55} with the relations \eqref{eq:hs56} is compatible with the flatness conditions \eqref{eq:hs7}.
It is worthwhile recalling that our variational principle requires $\tr\big(A_t\wedge\de A_x - \bar A_x\wedge\de\bar A_t\big)_{\partial\cal M}=0$.
Depending on the precise boundary conditions, to be determined by a canonical analysis, this may restrict some of the $f_n$ and $\bar a_n$ to be state-independent.

Let us fix $\hat r = \frac12\,\big(e^{\rho}+M\,e^{-\rho}\big)$, $a_1=M$, $\bar a_1=5f^2 M^2/32$, $f_{-1}=-M/(4f)$, $f_0=0$, $f_1=f$, $\bar f_{-1}=1/(4f)$, $\bar f_0 =0$ and $\bar f_1=-M f$.
The solutions are parametrized by two constants $M$ and $f$. 
The latter determines both $f_1$ and $\bar a_{-3}$.
For concreteness we choose $f=4$ [other values of $f$ would change only the most subleading term in the large $\hat r$ expansion of the line element \eqref{eq:hs57} below].
The line element constructed from the connections \eqref{eq:hs56} then describes asymptotic Lifshitz black holes with scaling exponent $z=3$ (and $\ell=\sqrt{10}$).
\eq{
\extd s^2 = \ell^2\,\Big[\frac{\extd\hat r^2}{\hat r^2-M} + \hat r^2\,\extd x^2 - \hat r^6\,\Big(1-\frac{3M}{2\hat r^2}+\frac{3M^2}{5\hat r^4}-\frac{49M^3}{400\hat r^6}\Big)\,\extd t^2\Big]
}{eq:hs57}
The curvature invariants constructed from the line element \eqref{eq:hs57} coincide with the ones constructed from the Lifshitz line element \eqref{eq:hs47} for $z=3$ in the limit $\hat r\to\infty$, but not for finite values of $\hat r$.
The geometry described by the line element \eqref{eq:hs57} has exactly one Killing horizon at $\hat r\approx 1.02 M$.

\section{Outlook}\label{sec:6}

In summary, the principal embedding of generic spin-$n$ gravity is tailor-made for the construction of asymptotic Schr\"odinger and Lifshitz spacetimes with a number of possibilities for the scaling exponent $z$ that grows quadratically with $n$.
The principal embedding never contains a singlet and is thus not suitable for the construction of asymptotically warped AdS or AdS$_2\times\mathbb{R}$ spacetimes. 
On the other hand, the non-principal embeddings typically contain one or more singlets and thus allow for the construction of either warped AdS or AdS$_2\times\mathbb{R}$ spacetimes (or, in many cases, both).
Higher spin gravity therefore provides a rich landscape of gravity duals to 2-dimensional quantum field theories in the context of the gauge/gravity correspondence beyond the canonical AdS/CFT holography.

We have provided the first steps towards non-AdS holography in 3-dimensional higher spin gravity.
We list now some of the next steps to be implemented in future work, as well as potentially interesting elaborations and generalizations.

The canonical charges presented in \cite{Banados:1994tn,Banados:1998gg} should be evaluated.
This will establish which boundary conditions to impose to guarantee finiteness, integrability and conservation of the canonical charges.
Moreover, evaluating their canonical brackets determines the central charges appearing in the asymptotic symmetry algebra.
Once the precise boundary conditions are known it will be interesting to check which black hole solutions are allowed in a given theory, similar to our discussion of Lifshitz black holes in section \ref{sec:5.1}.
Besides the BTZ black hole \cite{Banados:1992wn,Banados:1992gq}, which is locally and asymptotically AdS, there are several non-AdS black holes that have emerged in the literature on topologically massive gravity \cite{Deser:1982vy,Deser:1982wh} and new massive gravity \cite{Bergshoeff:2009hq,Bergshoeff:2009aq}, e.g.~warped AdS black holes \cite{Anninos:2008fx}, Schr\"odinger black holes \cite{Anninos:2010pm} and Lifshitz black holes \cite{AyonBeato:2009nh}.
Further checks of non-AdS holography are then possible, for instance the calculation of correlators on the gravity side together with a comparison with corresponding correlators on the field theory side or the calculation of 1-loop partition functions.
Thermodynamical considerations, including a microstate counting {\'a} la Cardy \cite{Strominger:1997eq,Birmingham:1998jt}, could provide a further piece of valuable information. 

Another important issue to be resolved is the (non-)uniqueness of the metric definition \eqref{eq:hs13} in terms of the zuvielbein for embeddings other than the principal one \cite{Campoleoni:2011hg,Castro:2011fm}.
In particular, since the singlets are $sl(2)$-invariant one can add any linear combination of them to the connections $A, \bar A$ in \eqref{eq:hs11}.
It would also be interesting to clarify which geometrical properties are actually gauge-invariant under the full gauge group, such as the causal structure, the asymptotic structure, the number and types of Killing horizons etc.

To investigate whether backgrounds are possible beyond the ones considered in the present work it would be of interest to classify and construct all stationary axi-symmetric solutions of generic higher spin gravity.
Technically, this is probably done most easily by exploiting Cl{\'e}ment's Ansatz \cite{Clement:1994sb} and to proceed along the lines of \cite{Ertl:2010dh}, where a complete classification of local stationary axi-symmetric solutions of topologically massive gravity was performed.
Alternatively, it seems auspicious to perform a Kaluza--Klein reduction analogous to \cite{Guralnik:2003we,Grumiller:2003ad}, which probably leads to a 2-dimensional dilaton gravity theory coupled non-trivially to some non-abelian gauge field(s). 
It might be rewarding to consider more general solutions than \eqref{eq:hs7} to the flatness conditions and to check whether they lead to more general asymptotic backgrounds than the ones considered here.
Concerning Lifshitz backgrounds it was pointed out recently that these spacetimes have a null curvature singularity \cite{Hartnoll:2009sz,Kachru:2008yh,Copsey:2010ya,Horowitz:2011gh}.
It would be interesting to check if these singularities are resolved in higher spin gravity, analog to the discussion in \cite{Castro:2011fm}.
Finally, it could be worthwhile to generalize our construction of non-AdS backgrounds to higher spin topologically massive gravity \cite{Chen:2011vp,Bagchi:2011vr,Chen:2011yx} and related higher spin massive gravity theories in three dimensions.

\acknowledgments

We thank Hamid Afshar, Andrea Campoleoni, Alejandra Castro, Niklas Johansson and Per Kraus for discussions.
DG is grateful to Sabine Ertl and Niklas Johansson for collaboration on warped AdS holography in topologically massive gravity.

MG and DG were supported by the START project Y435-N16 of the Austrian Science Fund (FWF). 
RR was supported by the FWF project P-22000 N-16 and the project DO 02-257 of the Bulgarian National Science Foundation (NSFB).

\appendix

\section{Suitable spin-4 bases}\label{app:A}

We use the following conventions for the $sl(2)$ generators
\eq{
[L_n,\,L_m] = (n-m) L_{n+m}
}{eq:app0}
where $L_{\pm 1}:= L_\pm$.
The remaining generators of the $W$-algebras obey the following commutation relations with the $sl(2)$ generators:
\eq{
[L_n,\,W_m^{l[a]}] = (nl-m)W_{n+m}^{l[a]} 
}{eq:app6}
The traces of these generators are given by
\eq{
\tr\big(W_m^{k[a]}\,W_n^{l[b]}\big) = (-1)^{l-m}\,\frac{(l+m)!(l-m)!}{(2l)!}\,\delta^{k,\,l}\,\delta_{m+n,\,0}\,N_l^{a,\,b}
}{eq:app7}
with the normalization
\eq{
N_l^{a,\,b}:=\tr\big(W_l^{l[a]}\,W_{-l}^{l[b]}\big)\,.
}{eq:app8}
Whenever singlets fall into an $sl(2)$ representation on their own we define their generators such that they obey the following commutation relations:
\eq{
[S^{[n]},\,S^{[m]}]=(n-m)S^{[n+m]}\,.
}{eq:app13}
We use the notation $S^{[n]}:=W_0^{0[n]}$.
Otherwise a singlet is denoted by $S$ without any index.
Doublets are denoted by $G_n^{[a]}$ (with $n=\pm$) so that $G_n^{[a]} := W_n^{1/2[a]}$.
Triplets are denoted by $T_n^{[a]}$ (with $n=0,\pm$) so that $T_n^{[a]} := W_n^{1[a]}$.
Quintets are denoted by $W_n$ (with $n=0,\pm 1, \pm 2$) so that $W_n:= W_n^{2[1]}$.
Septets are denoted by $U_n$ (with $n=0,\pm 1, \pm 2, \pm 3$) so that $U_n:= W_n^{3[1]}$.

\subsection{Principal embedding}\label{app:A.1}

Note that our conventions differ slightly from the ones by Tan \cite{Tan:2011tj}, who also provides a basis of generators for the principal embedding case of spin-4 gravity.\\
{\small 
$sl(2)$ generators:
\eq{
L_0 = \frac12\,\begin{pmatrix}3 & 0 & 0 & 0 \\ 0 & 1 & 0 & 0 \\ 0 & 0 & -1 & 0  \\ 0 & 0 & 0 & -3\end{pmatrix}\qquad
L_+ = \begin{pmatrix}0 & 0 & 0 & 0 \\ 3 & 0 & 0 & 0 \\ 0 & 4 & 0 & 0  \\ 0 & 0 & 3 & 0\end{pmatrix}\qquad
L_- = \begin{pmatrix}0 & -1 & 0 & 0 \\ 0 & 0 & -1 & 0 \\ 0 & 0 & 0 & -1  \\ 0 & 0 & 0 & 0\end{pmatrix}
}{eq:app20}
Quintet:
\eq{
\begin{split}
W_2 = 12\,\begin{pmatrix}0 & 0 & 0 & 0 \\ 0 & 0 & 0 & 0 \\ 1 & 0 & 0 & 0  \\ 0 & 1 & 0 & 0\end{pmatrix} \qquad
W_{-2} = \begin{pmatrix}0 & 0 & 1 & 0 \\ 0 & 0 & 0 & 1 \\ 0 & 0 & 0 & 0  \\ 0 & 0 & 0 & 0\end{pmatrix} \\
W_0 = \begin{pmatrix}1 & 0 & 0 & 0 \\ 0 & -1 & 0 & 0 \\ 0 & 0 & -1 & 0  \\ 0 & 0 & 0 & 1\end{pmatrix} \qquad
W_1 = 3\,\begin{pmatrix}0 & 0 & 0 & 0 \\ 1 & 0 & 0 & 0 \\ 0 & 0 & 0 & 0  \\ 0 & 0 & -1 & 0\end{pmatrix} \qquad
W_{-1} = \begin{pmatrix}0 & -1 & 0 & 0 \\ 0 & 0 & 0 & 0 \\ 0 & 0 & 0 & 1  \\ 0 & 0 & 0 & 0\end{pmatrix} \\
\end{split}
}{eq:app21}
Septet:
\eq{
\begin{split}
U_3 &= 36\,\begin{pmatrix}0 & 0 & 0 & 0 \\ 0 & 0 & 0 & 0 \\ 0 & 0 & 0 & 0  \\ -1 & 0 & 0 & 0\end{pmatrix} \;
U_2 = 6\,\begin{pmatrix}0 & 0 & 0 & 0 \\ 0 & 0 & 0 & 0 \\ -1 & 0 & 0 & 0  \\ 0 & 1 & 0 & 0\end{pmatrix} \;
U_{-2} = \frac12\,\begin{pmatrix}0 & 0 & -1 & 0 \\ 0 & 0 & 0 & 1 \\ 0 & 0 & 0 & 0  \\ 0 & 0 & 0 & 0\end{pmatrix} \;
U_{-3} = \begin{pmatrix}0 & 0 & 0 & 1 \\ 0 & 0 & 0 & 0 \\ 0 & 0 & 0 & 0  \\ 0 & 0 & 0 & 0\end{pmatrix} \\
U_0 &= \frac{1}{10}\,\begin{pmatrix}-3 & 0 & 0 & 0 \\ 0 & 9 & 0 & 0 \\ 0 & 0 & -9 & 0  \\ 0 & 0 & 0 & 3\end{pmatrix} \;
U_1 = \frac15\,\begin{pmatrix}0 & 0 & 0 & 0 \\ -6 & 0 & 0 & 0 \\ 0 & 12 & 0 & 0  \\ 0 & 0 & -6 & 0\end{pmatrix} \;
U_{-1} = \frac15\,\begin{pmatrix}0 & 2 & 0 & 0 \\ 0 & 0 & -3 & 0 \\ 0 & 0 & 0 & 2  \\ 0 & 0 & 0 & 0\end{pmatrix} \\
\end{split}
}{eq:app22}
}
Non-vanishing traces:
\begin{subequations}
\label{eq:app26}
\begin{align}
 \tr L_0^2 &= 5 & \tr\big(L_+ L_-\big) &= -10 & \tr\big(U_3 U_{-3}\big) &= -36 \\
 \tr W_0^2 &= 4 & \tr\big(W_1 W_{-1}\big) &= -6 & \tr\big(W_2 W_{-2}\big) &= 24 \\
 \tr U_0^2 &= \frac95 & \tr\big(U_1 U_{-1}\big) &= -\frac{12}{5} & \tr\big(U_2 U_{-2}\big) &= 6
\end{align}
\end{subequations}

\subsection{2-2 embedding}\label{app:A.2}

{\small
$sl(2)$ generators:
\eq{
L_0 = \frac12\,\begin{pmatrix}1 & 0 & 0 & 0 \\ 0 & 1 & 0 & 0 \\ 0 & 0 & -1 & 0  \\ 0 & 0 & 0 & -1\end{pmatrix}\qquad
L_+ = \begin{pmatrix}0 & 0 & 0 & 0 \\ 0 & 0 & 0 & 0 \\ 1 & 0 & 0 & 0  \\ 0 & 1 & 0 & 0\end{pmatrix}\qquad
L_- = \begin{pmatrix}0 & 0 & -1 & 0 \\ 0 & 0 & 0 & -1 \\ 0 & 0 & 0 & 0  \\ 0 & 0 & 0 & 0\end{pmatrix}
}{eq:app10}
Other triplets:
\begin{subequations}
\begin{align}
T_0^{[1]} &= \frac12\,\begin{pmatrix}1 & 0 & 0 & 0 \\ 0 & 0 & 0 & 0 \\ 0 & 0 & -1 & 0  \\ 0 & 0 & 0 & 0\end{pmatrix} &
T_+^{[1]} &= \begin{pmatrix}0 & 0 & 0 & 0 \\ 0 & 0 & 0 & 0 \\ 1 & 0 & 0 & 0  \\ 0 & 0 & 0 & 0\end{pmatrix} &
T_-^{[1]} &= \begin{pmatrix}0 & 0 & -1 & 0 \\ 0 & 0 & 0 & 0 \\ 0 & 0 & 0 & 0  \\ 0 & 0 & 0 & 0\end{pmatrix} \\
T_0^{[2]} &= \frac12\,\begin{pmatrix}1 & 1 & 0 & 0 \\ 0 & 0 & 0 & 0 \\ 0 & 0 & -1 & -1  \\ 0 & 0 & 0 & 0\end{pmatrix} &
T_+^{[2]} &= \begin{pmatrix}0 & 0 & 0 & 0 \\ 0 & 0 & 0 & 0 \\ 1 & 1 & 0 & 0  \\ 0 & 0 & 0 & 0\end{pmatrix} &
T_-^{[2]} &= \begin{pmatrix}0 & 0 & -1 & -1 \\ 0 & 0 & 0 & 0 \\ 0 & 0 & 0 & 0  \\ 0 & 0 & 0 & 0\end{pmatrix} \\
T_0^{[3]} &= \frac12\,\begin{pmatrix}0 & 0 & 0 & 0 \\ 1 & 1 & 0 & 0 \\ 0 & 0 & 0 & 0  \\ 0 & 0 & -1 & -1\end{pmatrix} &
T_+^{[3]} &= \begin{pmatrix}0 & 0 & 0 & 0 \\ 0 & 0 & 0 & 0 \\ 0 & 0 & 0 & 0  \\ 1 & 1 & 0 & 0\end{pmatrix} &
T_-^{[3]} &= \begin{pmatrix}0 & 0 & 0 & 0 \\ 0 & 0 & -1 & -1 \\ 0 & 0 & 0 & 0  \\ 0 & 0 & 0 & 0\end{pmatrix} 
\end{align}
\label{eq:app11}
\end{subequations}
Singlets:
\eq{
S^{[0]} = \frac12\,\begin{pmatrix}1 & 0 & 0 & 0 \\ 0 & -1 & 0 & 0 \\ 0 & 0 & 1 & 0  \\ 0 & 0 & 0 & -1\end{pmatrix}\qquad
S^{[+]} = \begin{pmatrix}0 & 0 & 0 & 0 \\ 1 & 0 & 0 & 0 \\ 0 & 0 & 0 & 0 \\ 0 & 0 & 1 & 0\end{pmatrix}\qquad
S^{[-]} = \begin{pmatrix}0 & -1 & 0 & 0 \\ 0 & 0 & 0 & 0 \\ 0 & 0 & 0 & -1 \\ 0 & 0 & 0 & 0\end{pmatrix}
}{eq:app12}
}
Non-vanishing traces:
\begin{subequations}
\label{eq:app25}
\begin{align}
 \tr L_0^2 &= 1 & \tr\big(L_+ L_-\big) &= -2  \\
 \tr\big(L_0 T_0^{[i]}\big) &= \frac12 & \tr\big(L_\pm T_\mp^{[i]}\big) &= -1 \\
 \tr\big(T_0^{[i]}T_0^{[j]}\big) &= \frac12\,M_{i,j} & \tr\big(T_+^{[i]}T_-^{[j]}\big) &= - M_{i,j}\\
 \tr S^{[0]\,2} &= 1 & \tr\big(S^{[+]} S^{[-]}\big) &= -2 
\end{align}
\end{subequations}
with $M_{i,j} := 1-\de_{i,1}\,\de_{j,3}-\de_{i,3}\,\de_{j,1}$.

\subsection{3-1 embedding}\label{app:A.3}

{\small
$sl(2)$ generators:
\eq{
L_0 = \begin{pmatrix}1 & 0 & 0 & 0 \\ 0 & 0 & 0 & 0 \\ 0 & 0 & 0 & 0  \\ 0 & 0 & 0 & -1\end{pmatrix}\qquad
L_+ = \sqrt{2}\,\begin{pmatrix}0 & 0 & 0 & 0 \\ 0 & 0 & 0 & 0 \\ 1 & 0 & 0 & 0  \\ 0 & 0 & 1 & 0\end{pmatrix}\qquad
L_- = \sqrt{2}\,\begin{pmatrix}0 & 0 & -1 & 0 \\ 0 & 0 & 0 & 0 \\ 0 & 0 & 0 & -1  \\ 0 & 0 & 0 & 0\end{pmatrix}
}{eq:app15}
Other triplets:
\begin{subequations}
\begin{align}
T_0^{[1]} &= \begin{pmatrix}0 & 0 & 0 & 0 \\ 0 & 0 & \frac{1}{\sqrt{2}} & 0 \\ 0 & 0 & 0 & 0  \\ 0 & 0 & 0 & 0\end{pmatrix} &
T_+^{[1]} &= \begin{pmatrix}0 & 0 & 0 & 0 \\ -1 & 0 & 0 & 0 \\ 0 & 0 & 0 & 0  \\ 0 & 0 & 0 & 0\end{pmatrix} &
T_-^{[1]} &= \begin{pmatrix}0 & 0 & 0 & 0 \\ 0 & 0 & 0 & -1 \\ 0 & 0 & 0 & 0  \\ 0 & 0 & 0 & 0\end{pmatrix} \\
T_0^{[2]} &= \begin{pmatrix}0 & 0 & 0 & 0 \\ 0 & 0 & 0 & 0 \\ 0 & -\frac{1}{\sqrt{2}} & 0 & 0  \\ 0 & 0 & 0 & 0\end{pmatrix} &
T_+^{[2]} &= \begin{pmatrix}0 & 0 & 0 & 0 \\ 0 & 0 & 0 & 0 \\ 0 & 0 & 0 & 0  \\ 0 & -1 & 0 & 0\end{pmatrix} &
T_-^{[2]} &= \begin{pmatrix}0 & -1 & 0 & 0 \\ 0 & 0 & 0 & 0 \\ 0 & 0 & 0 & 0  \\ 0 & 0 & 0 & 0\end{pmatrix} 
\end{align}
\label{eq:app16}
\end{subequations}
Quintet:
\eq{
\begin{split}
W_2 = & \begin{pmatrix}0 & 0 & 0 & 0 \\ 0 & 0 & 0 & 0 \\ 0 & 0 & 0 & 0  \\ -1 & 0 & 0 & 0\end{pmatrix} \qquad
W_{-2} = \begin{pmatrix}0 & 0 & 0 & -1 \\ 0 & 0 & 0 & 0 \\ 0 & 0 & 0 & 0  \\ 0 & 0 & 0 & 0\end{pmatrix} \\
W_0 = \frac16\, \begin{pmatrix}-1 & 0 & 0 & 0 \\ 0 & 0 & 0 & 0 \\ 0 & 0 & 2 & 0  \\ 0 & 0 & 0 & -1\end{pmatrix} \qquad
W_1 = \frac{1}{2\sqrt{2}}\, & \begin{pmatrix}0 & 0 & 0 & 0 \\ 0 & 0 & 0 & 0 \\ -1 & 0 & 0 & 0  \\ 0 & 0 & 1 & 0\end{pmatrix} \qquad
W_{-1} = \frac{1}{2\sqrt{2}}\,\begin{pmatrix}0 & 0 & 1 & 0 \\ 0 & 0 & 0 & 0 \\ 0 & 0 & 0 & -1  \\ 0 & 0 & 0 & 0\end{pmatrix} \\
\end{split}
}{eq:app17}
Singlet:
\eq{
S = \frac{1}{2\sqrt{3}}\,\begin{pmatrix}1 & 0 & 0 & 0 \\ 0 & -3 & 0 & 0 \\ 0 & 0 & 1 & 0  \\ 0 & 0 & 0 & 1\end{pmatrix}
}{eq:app18}
}
Non-vanishing traces:
\begin{subequations}
\label{eq:app24}
\begin{align}
 \tr L_0^2 &= 2 & \tr\big(L_+ L_-\big) &= -4 & \tr S^2 &= 1 \\
 \tr\big(T_0^{[1]}T_0^{[2]}\big) &= -\frac12 & \tr\big(T_+^{[1]}T_-^{[2]}\big) &= 1 & \tr\big(T_+^{[2]}T_-^{[1]}\big) &= 1 \\
 \tr W_0^2 &= \frac16 & \tr\big(W_1 W_{-1}\big) &= -\frac14 & \tr\big(W_2 W_{-2}\big) &= 1
\end{align}
\end{subequations}

\subsection{2-1-1 embedding}\label{app:A.4}

{\small
$sl(2)$ generators:
\eq{
L_0 = \frac12\,\begin{pmatrix}1 & 0 & 0 & 0 \\ 0 & 0 & 0 & 0 \\ 0 & 0 & 0 & 0  \\ 0 & 0 & 0 & -1\end{pmatrix}\qquad
L_+ = \begin{pmatrix}0 & 0 & 0 & 0 \\ 0 & 0 & 0 & 0 \\ 0 & 0 & 0 & 0  \\ 1 & 0 & 0 & 0\end{pmatrix}\qquad
L_- = \begin{pmatrix}0 & 0 & 0 & -1 \\ 0 & 0 & 0 & 0 \\ 0 & 0 & 0 & 0  \\ 0 & 0 & 0 & 0\end{pmatrix}
}{eq:app2}
Doublets:
\begin{align}
G_+^{[1]} &= \begin{pmatrix}0 & 0 & 0 & 0 \\ 0 & 0 & 0 & 0 \\ 1 & 0 & 0 & 0  \\ 0 & 1 & 0 & 0\end{pmatrix} &
G_+^{[2]} &= \begin{pmatrix}0 & 0 & 0 & 0 \\ 1 & 0 & 0 & 0 \\ 0 & 0 & 0 & 0  \\ 0 & 0 & 1 & 0\end{pmatrix} &
G_+^{[3]} &= \begin{pmatrix}0 & 0 & 0 & 0 \\ 1 & 0 & 0 & 0 \\ 0 & 0 & 0 & 0  \\ 0 & 0 & -1 & 0\end{pmatrix} &
G_+^{[4]} &= \begin{pmatrix}0 & 0 & 0 & 0 \\ 0 & 0 & 0 & 0 \\ 1 & 0 & 0 & 0  \\ 0 & -1 & 0 & 0\end{pmatrix} \nonumber \\
G_-^{[1]} &= \begin{pmatrix}0 & 1 & 0 & 0 \\ 0 & 0 & 0 & 0 \\ 0 & 0 & 0 & -1  \\ 0 & 0 & 0 & 0\end{pmatrix} &
G_-^{[2]} &= \begin{pmatrix}0 & 0 & 1 & 0 \\ 0 & 0 & 0 & -1 \\ 0 & 0 & 0 & 0  \\ 0 & 0 & 0 & 0\end{pmatrix} &
G_-^{[3]} &= \begin{pmatrix}0 & 0 & -1 & 0 \\ 0 & 0 & 0 & -1 \\ 0 & 0 & 0 & 0  \\ 0 & 0 & 0 & 0\end{pmatrix} &
G_-^{[4]} &= \begin{pmatrix}0 & -1 & 0 & 0 \\ 0 & 0 & 0 & 0 \\ 0 & 0 & 0 & -1  \\ 0 & 0 & 0 & 0\end{pmatrix} 
\label{eq:app3}
\end{align}
Singlets:
\eq{
S^{[0]} = \frac12\,\begin{pmatrix}0 & 0 & 0 & 0 \\ 0 & 1 & 0 & 0 \\ 0 & 0 & -1 & 0  \\ 0 & 0 & 0 & 0\end{pmatrix}\qquad
S^{[+]} = \begin{pmatrix}0 & 0 & 0 & 0 \\ 0 & 0 & 0 & 0 \\ 0 & 1 & 0 & 0  \\ 0 & 0 & 0 & 0\end{pmatrix}\qquad
S^{[-]} = \begin{pmatrix}0 & 0 & 0 & 0 \\ 0 & 0 & -1 & 0 \\ 0 & 0 & 0 & 0  \\ 0 & 0 & 0 & 0\end{pmatrix}
}{eq:app4}
\eq{
S = \begin{pmatrix}1 & 0 & 0 & 0 \\ 0 & -1 & 0 & 0 \\ 0 & 0 & -1 & 0  \\ 0 & 0 & 0 & 1\end{pmatrix}
}{eq:app5}
}
Non-vanishing traces:
\begin{subequations}
\label{eq:app23}
\begin{align}
 \tr L_0^2 &= \frac12 & \tr\big(L_+ L_-\big) &= -1 & \tr\big(G_+^{[i]} G_-^{[j]}\big) &=  2\,\de_{i-2,j} -2\,\de_{i,j-2}  \\
 \tr S_0^2 &= \frac12 & \tr\big(S_+ S_-\big) &= -1 & \tr S^2 &= 4 
\end{align}
\end{subequations}


\providecommand{\href}[2]{#2}\begingroup\raggedright\endgroup

\end{document}